# Shared User Interfaces of Physiological Data: Systematic Review of Social Biofeedback Systems and Contexts in HCI

## Clara Moge[1], Katherine Wang[1], and Youngjun Cho[1,2*]

[1] UCL Interaction Centre (UCLIC), University College London, London, United Kingdom

[2] Department of Computer Science & WHO Collaborating Centre for Research on Assistive Technology, University College London, London, United Kingdom

## ABSTRACT

As an emerging interaction paradigm, physiological computing is increasingly being used to both measure and feed back information about our internal psychophysiological states. While most applications of physiological computing are designed for individual use, recent research has explored how biofeedback can be socially shared between multiple users to augment human-human communication. Reflecting on the empirical progress in this area of study, this paper presents a systematic review of 64 studies to characterize the interaction contexts and effects of social biofeedback systems. Our findings highlight the importance of physio-temporal and social contextual factors surrounding physiological data sharing as well as how it can promote social-emotional competences on three different levels: intrapersonal, interpersonal, and task-focused. We also present the Social Biofeedback Interactions framework to articulate the current physiological-social interaction space. We use this to frame our discussion of the implications and ethical considerations for future research and design of social biofeedback interfaces.

## 1 INTRODUCTION

Designing computing systems able to sense, analyze, and respond to emotional states is essential in bringing human-computer interaction closer to human-human communication [102]. These physiological and affective computing technologies (also called emotional artificial intelligence) rely upon input data from .,physiological signals such as blood volume pulse, heart rate variability, temperature and respiration [18, 21, 22, 122, 139], as well as behavioral cues (e.g. facial expression, gesture, and tone [103, 134]). Through the process of psychophysiological inference (i.e., the mapping of phenomenologically distinct psychological states to patterns of physiological activity [19]), these inputs are analyzed to interpret cognitive, emotional and motivational states. This allows the computing agent to modulate its response as a function of user state to drive task efficiency and improve user experiences [36]. On the other hand, instead of modulating its response, the computer can also return real-time physiological information to users in a process known as biofeedback [39]. This involves externalizing physiological state in an accessible form for users to monitor and subsequently learn to control their own physiological activity [39, 139]. For this reason, biofeedback-based physiological systems have been particularly beneficial in therapeutic contexts, including in anxiety disorder management [92], substance abuse treatment [115] and general stress management [139], and have also gained popularity in health and fitness self-tracking [96].

* youngjun.cho@ucl.ac.uk

Recently, physiological, affective, and social computing research agendas have converged to consider the question: *if biofeedback can be used on an intrapersonal level to help people understand their psychophysiology, can it be used interpersonally to promote an understanding of others?* This question is fueled by studies showing physiological signals are inherently emotional *and* social [16]. For example, physiological interaction between pairs is correlated with attachment, engagement, and team performance (see [97] for a review), and individual physiology reflects changes in social emotions like embarrassment [51], guilt [25] and empathy [28] as well as attention [131] in various interpersonal contexts. While physiological activity reflects our reactions to social interactions [131] and can inform us about the emotional states of others [111], physiological signals do not constitute an explicit channel of information that humans use to interpret others' states [16] as most signals are subtle and displayed internally [16]. However, some research has argued that externalizing such data through biofeedback in social contexts can help individuals recognize their own and others' emotions [16, 76], and in turn augment social communication [76]. This shift from using physiological input to promote human-computer and intrapersonal interactions to now enhancing human-human interaction has significant implications in fulfilling people's need to belong [4]. For the remainder of this paper, we refer to technologies enabling the sharing of physiological signals as *social biofeedback* systems.

While it is promising, there are limitations to our understanding of social biofeedback, its effects and the environmental factors surrounding its use. A reconceptualization of physiological signals as a communication medium has been proposed [38]; however, there is no systematic understanding of the contexts of social biofeedback systems. This is important given the significant moderating influence that context can have on the quality of communication [9] and thus social outcomes. To address this, we systematically review the existing literature on social biofeedback systems to understand how these technologies can enrich social interaction. We synthesize the physical, temporal, and social contexts in which such systems are embedded and articulate the current physiological-social space through a novel framework, the Social Biofeedback Interactions framework. We also provide a qualitative analysis of the effects of social biofeedback on socio-emotional skills, and a meta-analysis of its effects on positive emotions.

## 2 LITERATURE REVIEW

### 2.1 Situating physiological computing in a social context

Biofeedback-based computing systems are increasingly being embedded into everyday ambulatory contexts [139] which naturally involve the presence of other people. However, of specific interest to this review is how biofeedback can be *intentionally* designed to promote outcomes of social interaction. This idea is highlighted in Chanel and Mühl 's seminal paper [16], which outlines two principal directions for physiological computing in social interactions: (1) using physiology as a social cue and (2) using physiology as a measurement of social interactions. As physiological data carries socially relevant information, allowing others to perceive it can help people make mental inferences, regardless of whether psychological synchrony is involved [127]. The asymmetry involved in user-observer interactions has also been useful in assistive contexts, specifically to tailor communication to support educational (e.g., [138]) and work-related outcomes (e.g., [126]). Additionally, since physiological activity fluctuates in response to social factors, biofeedback systems can also be used to assess group-level social processes like collaboration. This can be done where the computing module of the system assesses the interaction (e.g., synchrony) between physiological inputs from multiple users [16].

To the best of our knowledge, only one systematic review has since explored the landscape of social biofeedback technologies [38]. This review conceptualized physiological data sharing as a novel communication medium from a media psychology and communication perspective. By assessing the communication characteristics of physiological data sharing, they found that while most current social biofeedback systems enable communication in real-time, they do not afford much autonomy nor opportunities to revise input before sharing it with others. Potentially reflecting trends toward interactional (as opposed to informational) displays of biofeedback in HCI [6], researchers also found many systems afforded reciprocity (bi- or multi-directional data sharing). This is interesting because the support of multi-user input shows a new physiologically-mediated mode of communication [16].

It was also highlighted that sensemaking with displays of physiological data is highly context-dependent, with notable contextual influences being the way data is visualized and the relationship between interactional parties [38]. This is consistent with Bradley and Dunlop's [9] model of contextual information in computer science and HCI, which considers both the computing and social environments. The types of relationships between users, which fall under the social context category, have not yet been studied alongside the physical and temporal dimensions of context. Characterizing social biofeedback systems based on these moderating factors surrounding their contexts of cases is important to understand the mechanisms through which these systems lead to beneficial effects.

## 2.2 Towards an understanding of transferable benefits of social biofeedback

In their systematic review, Feijt et al. [38] found that social biofeedback systems can benefit interpersonal relationships through increasing feelings of intimacy, connectedness and shared experiences. This is consistent with other research showing that social connectedness can be enhanced by adding modality-specific features designed to express emotions like facial expressions, gestures and emoticons [59] (e.g., Hug Over a Distance [128], the Sensing Beds [48], and the Cube and the Picture Frame [43]).

While results from [38] shed light on the benefits of social biofeedback from a user experience (UX) standpoint, what remains unclear is whether these effects are transitory. We turn to the development of social-emotional skills as a potentially enduring and transferrable benefit of social biofeedback as they enable people to express, regulate and understand their affective, cognitive and behavioral states in everyday life and social interactions [114]. Interestingly, studies have focused on exploring biofeedback systems to specifically support processes of individual emotional expression, regulation and understanding [35]. However, there have not been considerations of shared user interface perspectives, particularly amidst concerns that excessive screen-time and Internet use may hinder social and emotional skill development [105].

## 2.3 Problem statement and contributions

While the communication characteristics of biofeedback have been explored [38], no studies have systematically identified the contexts of social biofeedback systems. It is also unclear whether such systems can help users practice socio-emotional skills, beyond providing transitory experiences. Addressing these points, the contributions of this paper are: (1) a systematic understanding of the physio-temporal and social characteristics of the interaction contexts of social biofeedback systems, (2) the presentation of a novel framework, the Social Biofeedback Interactions framework, articulating the current physiological-social interaction space, (3) a qualitative analysis of socio-emotional competences involved in interacting through shared interfaces of physiological data, and (4) a meta-analytic review of the effectiveness of social biofeedback interactions on positive emotions.

## 3 METHOD

Our review is guided by two research questions: what types of interactions occur in physiological-social space, and with whom, and what kind of socio-emotional competencies can be practiced and/or developed by sharing displays of physiological data? Our method for identifying and screening research papers follows guidelines by the Preferred Reporting Items for Systematic Reviews and Meta-Analyses (PRISMA; see our detailed flowchart in Appendix A). We also conduct a thematic analysis and meta-analysis on the included studies.

### 3.1 Search strategy

The literature search was conducted between June-July 2021 in four databases: ACM Digital Library, IEEE Xplore, Scopus and SpringerLink. Google Scholar was also used as an additional resource to ensure all relevant papers were included and locate others potentially missed by the database search. References of select papers [16, 58, 119] which were key to the research questions were also checked for relevant papers.

The Boolean search string was iteratively developed across two concept categories, physiological computing and social interaction. The final search string used across full texts: ("physiological sens*" OR "physiological signal*" OR "physiological data" OR biosens* OR biosignal* OR biodata OR biofeedback) AND (social OR interpersonal). To narrow search results from the broader Scopus database, the following search filters were applied: journal and conference proceedings, written in English, within Psychology, Engineering or Computer Science categories, and with suggested keywords "human", "physiology" or "biofeedback". Similarly, the filters applied for the SpringerLink database were: articles in "Computer Science and User Interfaces" and "HCI" categories.

### 3.2 Screening process

The initial search returned a total of 3,951 research articles which imported into Zotero reference management software [1]. 1,233 duplicates were automatically deleted by the software and five additional duplicates were then deleted manually. The remaining articles (n= 2,713) were then screened for relevance which was assessed based on the inclusion of keywords present in the search string and a focus on computer-mediated human-human interaction. For instance, papers explicitly describing physiological signals only in monitoring systems and for evaluation of clinical/rehabilitative interventions were excluded at this stage.

### 3.3 Selection process and eligibility criteria

The screening resulted in 180 full-text research articles to assess for eligibility. For journal articles, conference proceedings and dissertations to be eligible for review, studies were required to (1) be published in English, (2) be peer-reviewed, (3) include at least one defined, indirect physiological measure, (4) involve human participants and (5) describe an experiment, prototype or system involving a minimum of two interactional partners. Studies investigating interpretations of *manipulated* physiological data (e.g., [30], [135]) and from *imagined* interactional partners (e.g., [89], [90]) were also included, as we were interested in the perception of physiological data in social contexts, even if hypothetical. Where authors published several papers on a given system or prototype, papers were included if considered as making a novel, directly relevant contribution. Otherwise, only the most comprehensive paper was included.

To further narrow the scope of the review, studies were excluded if they (1) described only physiological measurement by a third-party, such as using physiology for patient monitoring purposes or (2) used only external measures (e.g., facial expression, gesture), endocrine or behavioral measures of physiology. We exclude external measures as they represent affect in the physical rather than physiological domain [106]. Papers capturing a combination of direct and indirect measures (e.g., [109], [107]) were included. Those describing (3) a prototype or design concept without evaluation or (4) physiological computing applications only in assistive contexts like therapy (e.g., [68]), education (e.g., [138]) or occupation (e.g., [126]) were excluded. Papers were also excluded if they (5) described only individual uses of biofeedback contexts. We also included papers in which individual biofeedback was designed to be visible to others, such as in a public space (e.g., [64], [112]) or multiuser environment (e.g., [71]), as we considered such contexts to be social. Finally, papers were excluded if they (6) adopted a purely theoretical perspective, or (7) constituted research proposals, system demonstrations or doctoral consortiums.

The three authors developed the inclusion and exclusion criteria which the first author initially applied to the papers. To ensure inter-rater reliability, the second author verified screening decisions for a randomly selected subset of 20 papers, and the third author verified the complete sets of included and excluded papers. All three authors met regularly to define eligibility criteria, and the list of included papers was reviewed and assessed continually.

Leading causes of exclusion were the use of physiology for measurement and monitoring purposes (n= 23) and a lack of reporting of prototype evaluations (n= 21). The full text of 1 paper [101] was not available on any online database, despite the authors' best efforts to locate it, and was therefore excluded from the corpus. In addition, two papers ([12] and [108]) were excluded on the basis that more comprehensive papers describing identical prototypes were already included for review ([52] and [109], respectively). Finally, 72 papers were included.

### 3.4 Quality assessment

To assess risks of bias, we conducted a quality assessment using the Critical Appraisal Skills Programme (CASP) qualitative research checklist [14], the most recommended tool for qualitative studies on individual experiences and in social contexts [84]. The checklist was adapted for studies describing a quantitative or mixed-methods approach. Based on this assessment, papers were rated as having "good" (n= 46), "fair" (n= 12) or "poor" (n= 8) methodological quality. The initial assessment was done by the first author and verified by the remaining authors where the quality of a paper was deemed poor, predominantly due to a lack of rigorous data analysis. All papers receiving a "poor" rating were excluded, resulting in the final inclusion of 64 studies.

### 3.5 Data extraction

Key bibliometrics from each paper (e.g., title, author(s), publication year) were exported from Zotero into an Excel spreadsheet. We then manually extracted relevant data from each study: study participants, materials and prototypes, design and methodology, key outcomes of biofeedback sharing, and social effects reported.

For papers wherein multiple studies were reported, data was extracted for each relevant study reporting user feedback on a prototype or outcomes from an intervention with an implemented prototype. For example, data was not extracted for user studies conducted for preliminary user research (e.g., in [93]), pilot tests for general usability (e.g., in [88]) or specific prototype features (e.g., in [77]), and for physiological measurement validation or calibration before intervention (e.g., in [41]).

### 3.6 Thematic analysis

A thematic analysis was conducted to examine the effects of human-human interaction mediated by social biofeedback on socio-emotional skills. The aim was to qualitatively describe these effects across affective, cognitive, and behavioral levels. Following the process for deductive thematic analysis outlined by Braun and Clark [10], we started by familiarizing ourselves with the data and taking preliminary notes. The data consisted of the authors' interpretations of their primary data. Initial codes were then generated and iteratively sorted into higher-level themes, reviewed for validity, and further refined. In total, the thematic analysis was performed on 52 studies, after excluding those which reported only quantitative results (n=12).

### 3.7 Meta-analysis

A meta-analysis was conducted to estimate effect sizes for the effectiveness of social biofeedback interactions on positive affect. We included all studies reporting a quantitative measure of positive affect captured after social biofeedback-mediated interaction (n=6). Mean scores and standard deviations (SDs) of affective questionnaires measuring positive affect were extracted. Cohen's $d$ [24] was used to estimate the standardized mean difference (SMD) with 95% confidence intervals (CI). Inferential statistical test values (e.g., t-test values, F-values) were used to calculate effect size when means and standard deviations were not reported.

## 4 RESULTS

### 4.1 General study characteristics

Studies on social physiological data sharing included in this work were published between 2007-2021 (see Appendix B for details). The number of studies on the topic is gradually increasing, particularly since 2013, despite a potentially lower emergence of studies from 2020 due to feasibility issues during the COVID-19 pandemic. General methodology and design characteristics of each paper are detailed in Appendix C. Further details of physiological sensors, biofeedback modalities and forms explored in the studies can be found in Appendix D.

The most common application contexts for social biofeedback are social play and tele-social communication (both n=14, 22% each). Social play use cases range from multiplayer virtual reality (VR) gameplay (n=5, [29–31, 45, 63]) to video gaming (n=3, [32, 95, 107, 142]), tabletop gameplay (n=3, [2, 32, 40]), mobile (n=2, [71, 73]) and outdoor gameplay (n=1, [87]). Use cases for tele-social communication are primarily in remote non-verbal communication (n=9, [3, 41, 58, 65, 77, 91, 125, 133, 135]), followed by text-messaging (n=4, [52, 67, 80, 81]) and video chatting (n=1, [70]). Other prominent application contexts for social biofeedback include public interactive displays of emotion (n=11, 17%) in the form of installations (n=6), performances (n=3) and wearables (n=3), as well as mediated social interaction (n=7), social meditation and relaxation (n=6), social exertion (n=5), online content sharing (n=4) and face-to-face communication (n=2). Overall, the oldest applications of social biofeedback are in social entertainment, play and exertion, with studies emerging since 2007. Mostly from 2015 onwards, an increasing number of studies have focused on tele-social and augmented emotion communication, with exploration through artistic means in the last few years.

## 4.2 Characteristics of physio-temporal and social contexts

### 4.2.1 Physical-temporal context

To understand the physical and temporal characteristics of physiological data sharing in social contexts, we use the Time-Space taxonomy of groupware proposed by Ellis et al. [34], which distinguishes co-located vs. distributed physical locations and synchronous (real-time) vs. asynchronous temporal dimensions of interaction. Our analysis is shown in Figure 1.

| **Physical context** | | Synchronous | Asynchronous |
|---|---|---|---|
| | Distributed | [29], [3], [26], [27], [30], [31], [33], [41], [45], [58], [60], [61], [63], [65], [70], [85], [94], [95], [107], [109], [113], [82], [125], [135] | [52], [67], [73], [72], [80], [78], [75], [81], [77], [89], [90], [91], [118], [133], [136] |
| | Co-located | [2, 63, 93] , [17], [32], [40], [46], [54], [87], [88], [100], [110], [116], [121], [122], [123], [142] | [37], [55], [98], [112], [119], [120], [83], [130] |
| | | Synchronous | Asynchronous |

**Temporal context**

Figure 1: 2x2 matrix showing studies categorized according to the Time-Space Taxonomy of groupware [34].

Distributed systems in general (n=39, 61%) were more commonly described than co-located ones (n=25 while synchronous systems (n=51, 80%) were more common than asynchronous ones (n=23). Specifically, distributed synchronous social biofeedback systems constitute the biggest category (24 studies, 38%), followed by co-located synchronous (n=17, 27%), distributed asynchronous (n=15, 23%) and co-located asynchronous systems (n=8, 13%).

### 4.2.2 Articulating the physiological-social interaction space

Given the range of social contexts and relationships between interactional partners described in the literature, we considered how to represent the social interaction space for current biofeedback systems. First, driven by the concept of using one individual's biofeedback as a social cue for third-party observers [16], we formalize the notion of asymmetrical interaction in physiological-social space. While the participant-observer model of biofeedback has previously been described [31], we considered how to build on this to reflect differences in biofeedback access, extending the idea that in media space, making information (e.g., physiological state) available is independent of necessarily obtaining it [44]. We refer to access as opposed to visibility to encompass multi-modal forms of biofeedback. Hence, within asymmetrical interactions, we distinguish between systems that afford biofeedback access to both primary and secondary users (observers; Figure 2a) and those which only display biofeedback to secondary users (Figure 2b). As we intentionally excluded studies describing researchers as recipients of users' physiological data, we emphasize here that secondary user(s) refers to observers without an experimental agenda.

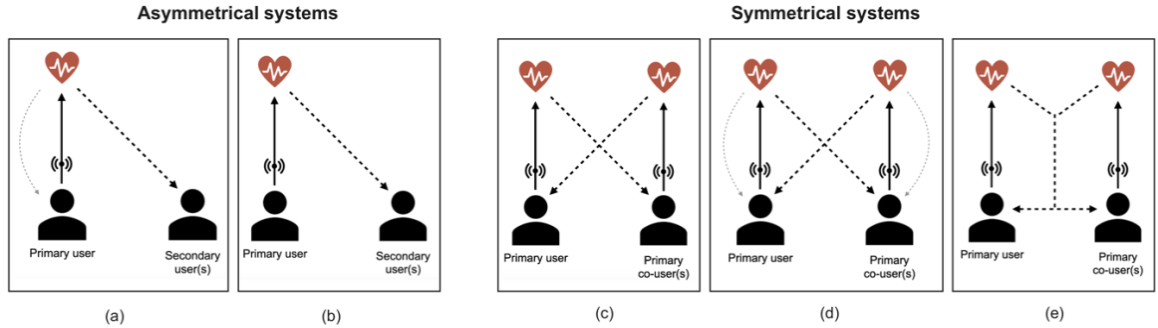

Figure 2. Social biofeedback data flows in (a, b) asymmetrical and (c - e) symmetrical systems. The heart icon represents any general indirect physiological activity. Solid black arrows show physiological input, curved grey arrows show biofeedback of self-data, and dotted arrows show social biofeedback.

In addition, we extend the notion of symmetry in media space [129] to describe a theoretically unexplored class of symmetrical interactions in physiological-social space where physiological input is obtained from more than one user. Unrelated to time, symmetry distinguishes itself from synchrony (e.g., text messaging can be both symmetrical and asynchronous). In practice, sharing physiology in multi-user environments goes beyond individual displays to what we refer to as *'collective'* representations of physiological data. Within collective displays, we also consider variations in *biofeedback content*, as opposed to access, since both primary users are given access in symmetrical communication systems. Biofeedback content can include data from the other user only (Figure 2c), or data from both users either in raw (Figure 2d) or aggregated forms (Figure 2e). The characteristics of symmetrical and asymmetrical systems are summarized in Table 1.

Table 1: Summary of discriminating characteristics of symmetrical and asymmetrical social biofeedback systems

| System | Representation | Biofeedback parameters | Users |
| --- | --- | --- | --- |
| Symmetrical | Collective | Content | Both primary |
| Asymmetrical | Individual | Access | Primary and secondary |

We articulate the full scope of the current physiological-social space in Figure 3: the *Social Biofeedback Interactions Framework*. The framework shows the basic physiological data flows and multi-user interactions around individual and collective physiological data displays. It also considers behavioral forms of feedback between users.

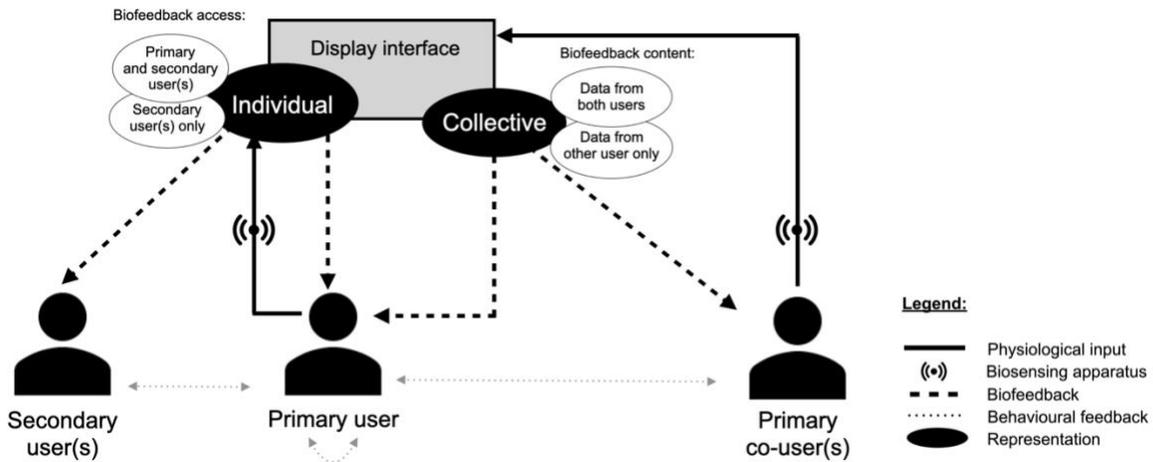

Figure 3. The Social Biofeedback Interactions Framework

Overall, 37 studies (58%) designed symmetrical social biofeedback systems, 23 studies (36%) designed asymmetrical ones and four studies engineered opportunities for both. In symmetrical systems, it was most common for users to receive biofeedback about their own physiological state together with that of other users' (n=14; [2, 52, 54, 67, 70, 72, 73, 75, 77, 85, 95, 116, 119, 121]). In 12 studies, users only received feedback about states of co-users and not their own [3, 29, 30, 45, 55, 65, 87, 91, 94, 110, 125, 133], while in four studies only aggregated measures of physiology were shown to both players [83, 100, 107, 142]. Six studies combined the sharing physiology of all users with additional aggregated measures like average metrics [17] and synchrony level between users [46, 60, 61, 82, 113]. One study manipulated both receiving biofeedback from self and others, instead of only others' [40].

In asymmetrical systems, it was most common for the primary user not to be in the loop, with only secondary users having access to the individual biofeedback (n=13, 54% of asymmetrical systems, [26, 27, 31, 41, 58, 78, 81, 88–90, 130, 135, 136]). Moreover, biofeedback access was granted to both primary and secondary users in 11 studies [37, 63, 64, 80, 93, 98, 109, 112, 118, 120]. Finally, the four studies designed for both symmetrical and asymmetrical interaction styles involved two or more symmetrically-engaged users and bystanders who could observe the interaction [32, 33, 122] and provided behavioral feedback [123].

### 4.2.3 Social communication context

Previous research has shown that the effects of physiological data sharing depend on the relationship between interacting individuals [119, 122]. To understand the roles of social biofeedback systems as a function of their symmetry affordances, we categorize prototypes with respect to four types of social communication contexts: interpersonal, group, public and mass [15] (see Figure 4). Within each, we identify different types of social relationships.

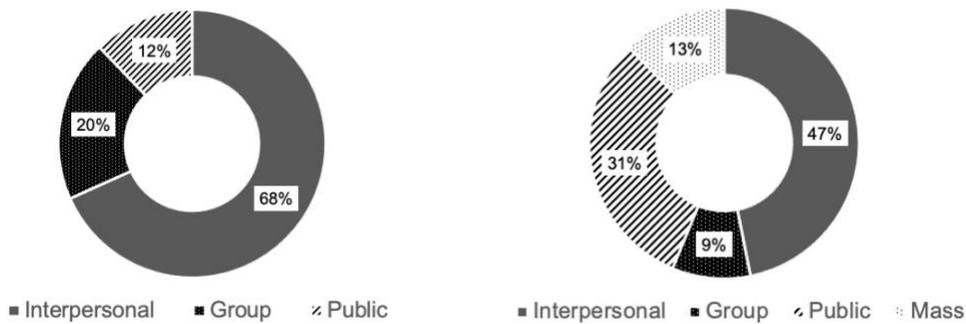

Figure 4: Social communication contexts for (left) symmetrical and (right) asymmetrical social biofeedback systems

Most symmetrical social biofeedback systems were geared towards connecting dyads (n=28) with strong attachments, such as close relations [3, 52, 54, 65, 67, 70, 75, 91, 116, 125] and romantic partners [65, 77, 110, 119, 133]. Such systems were also used to connect people with similar hobbies, like video gaming [45, 95, 107] and running [93]. Bidirectional sharing of physiological data in groups (n=8) was used mostly as a mechanic in social gameplay with friends [2, 32, 40, 87] and during shared physical exertion with fellow athletes [85, 121]. Symmetrical systems deployed in public contexts (n=5) focused on connecting members of the public by socially engaging audience members [100, 123] and designing opportunities for transformative interpersonal experiences with strangers [33, 55, 122].

Asymmetrical sharing in dyadic interaction was mainly designed for experimental purposes to study mechanisms of social perception [27, 31, 58, 78, 81, 89, 90, 135, 136], with one study using it to introduce strategic interdependence in two-player VR games [63]. In group contexts, two studies shared physiological metrics of leaders during group exercise [88, 93]. In public, such systems (n=10) involved externalizing emotional state as a form of self-expression with loved ones [41], friends [120] and colleagues [112], but also in general day-to-day social interactions with strangers [37, 98]. Asymmetrical biofeedback was also used to celebrate physical effort among athletes [64, 130]. Lastly, mass communication contexts involved one user sharing their physiological activity on social networking platforms, including live broadcasts during specific events [26, 80] as well as during video streaming [109] and sharing [118].

### 4.3 The effects of social biofeedback systems

Figure 5 shows the six themes synthesized from 61 codes through thematic analysis: (1) mindful self-awareness, (2) self-reflection and regulation of affective states, (3) empathy, (4) compassion and caregiving, (5) relationship skills for authentic connectedness, and (6) motivation, performance, and coordinative effort. These themes are also sorted according to the key conceptual domains of socio-emotional competencies outlined in the domains and manifestations of social-emotional competences (DOMASEC) model [114]. The framework delineates how individuals perceive themselves (self-orientation) from how they interact with people around them (others-orientation) and how they engage with tasks in the environment they are in (task-orientation). We use this multi-disciplinary framework as it allows us to clearly distinguish between affective, cognitive, and behavioral manifestations of the effects of social biofeedback systems.

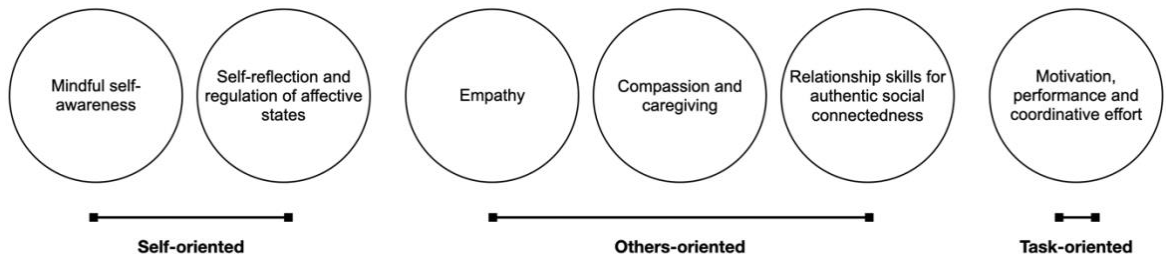

Figure 5: Overview of major themes and corresponding DOMASEC domains identified through thematic analysis.

### 4.3.1 Theme 1: Mindful self-awareness

As an intrapersonal skill, self-awareness describes the ability to recognize our emotions. In defining this theme, we associate this with mindfulness to describe non-reflective and non-judgmental awareness of our own affective states [13]. In this work, mindful self-awareness was found in meditative states of relaxation, noticing new internal and external phenomena, and attitudes of openness and acceptance.

Several studies found that sharing physiological data in immersive environments was a relaxing and calming experience [33, 41, 46, 52, 55, 120, 122, 125, 133]. While generally being positive in valence, these experiences were not associated with positive high arousal emotions (e.g., joy) but were instead associated with low-arousal states (e.g., pleasant, nice, and peaceful) [31, 41, 125]. Physiologically, peaceful states were found in trends towards decreasing arousal of users [125] and a natural synchronization of physiological activity [41, 122]. Some studies also found that users made fewer efforts to communicate with co-users in meditative states, instead becoming more quiet [31, 55, 120, 125]. In others, users aware of their own mindful self-awareness expressed wanting to help co-users enter the same state [33, 46]. The emergence of contemplative meditation also helped to elicit new perspectives on the self through feelings of individual vulnerability [55] and uniqueness [80]. Finally, where social biofeedback was explicitly used for meditative relaxation, studies found an increased sense of focus [46] and non-mindful hyper-awareness of one's weaknesses relative to other users [93]. The cognitive mechanisms behind this line of heightened awareness are described in Theme 2.

Some studies found forms of detached awareness allowing users to perceive internal bodily sensations better, noticing new relationships between their physical bodies and the environment [3, 71, 94]. The meta-awareness of new relationships was an interesting finding in a few studies, with links observed between body and mind [77], as well as between users and wider lifeforms such as nature [55, 122]. Some studies also found that users noticed relationships between emotions [116] and between users over time [55]. Studies also found that users were curious about learning more about themselves, consciously experimenting with their bodies [52, 64, 116, 119, 120].

Our analysis also revealed important attitudinal shifts towards openness and acceptance, both key to mindfulness practice [117]. Several studies found that the mere act of sharing physiological data was perceived as a gesture of openness [33, 71, 77, 80, 119]. Openness was also found in that some users made conscious efforts to observe both the objective display of their physiology and their subjective states, inviting the possibility of differences between the two [27, 54]. In terms of fostering acceptance, some studies found users embracing a loss of agency despite privacy concerns [64, 88, 116, 130], as well as accepting negative emotions as part of the process of physiological data sharing [116].

### 4.3.2 Theme 2: Self-reflection and regulation of affective states

On a cognitive level, self-reflection involves the active examination of one's beliefs, thoughts, and knowledge. Unlike mindful self-awareness, it describes the elaborative processing of experience [99].

Studies found that engaging with social biofeedback systems encouraged individual users to reflect on the specific causes of their emotions [37, 52, 54, 83, 98, 120], judge others' emotions [89, 90, 110] and identify triggers for negative states in particular [52, 112]. Shared physiological data displays also stimulated comparative judgments and reflected on one's own states relative to another user's [27, 120, 121]. After making judgements about their own affective states, biofeedback displays were used to validate these judgments as a form of confirmation [27, 54, 98, 100, 116]. Other studies found this was also the case when judgements were made about the states of others [75, 110].

These evaluations of personal emotional states also led to behavioral efforts to self-regulate. For instance, some studies found users censored or tailored their behavior to avoid potential tension with other users [37, 110, 112]. Behavioral self-regulation was not only found for those whose physiological data was displayed but also for secondary users [37]. Other studies found self-regulation occurring for impression management, where users consciously manipulated their self-presentation to deceive, for example, in games [32, 40, 46] and sometimes during communication with a partner [67, 80].

We found that reflection during interaction with shared displays of physiological data can also elicit worries about self-revelation. For example, studies found that users were concerned about potential discrepancies between their own appraisal of their emotional state and the objective display [37, 98, 112, 119], as well as between their own appraisal and that of other users [32, 33, 54, 75, 78, 80, 112, 116]. This was related to maladaptive cognitions such as inflated self-consciousness [93, 120] and preoccupations and fears of embarrassment [32, 33, 85].

### 4.3.3 Theme 3: Empathy

We found that both affective and cognitive forms of empathy were relevant competences used during human-human interaction in physiological-social space. Affective empathy describes the ability to share another person's emotional state, while cognitive empathy is the ability to accurately recognize the state of mind of another [13].

Several studies found affective empathy in the form of shared positive emotions like happiness, vitality and enthusiasm [3, 55, 64, 118, 123]. Positive emotions were also shared when users went through hardship together, for instance, in social physical exertion [94, 121, 130] or during games [63, 95]. Some studies found direct effects of one user's emotional state on another's, suggesting emotional convergence [63, 122, 125]. Others found evidence of emotional convergence where users tried to mimic a pattern of physiological response [41, 65, 67] or where physiological activity naturally became synchronized between interactional partners [52, 122, 123, 125]. Haptic 'feeling' was also found to directly increase feelings of empathy between users [72, 118]. Some studies also found evidence of empathic concern, where users became worried when another user's physiology was too high or low [45, 75, 130]. However, several studies also found that having access to another person's physiological data had the opposite effect of connecting people, such as alienation when physiology was too different [3, 93, 121].

We also found that cognitive empathy was fostered by having access to the physiological activity of another. Studies found that when presented with the biofeedback of another person, users were able to accurately detect others' internal states, a construct of empathy called mind perception [27, 37, 52, 65, 116, 125, 130]. Beyond perception, other studies found biofeedback of co-user was helpful to *understand* the emotional state of that person [31, 75, 78, 89, 90, 118]. In some cases, users went further by actually asking probing questions about the biofeedback to fully understand the context of their co-user [52, 75, 110, 120]. Finally, many studies also found users made guesses and explicit inferences about others' states, another sub-construct of empathy called mentalizing [40, 52, 54, 65, 75, 83, 89, 90, 110, 116, 119, 120].

Some mixed-methods studies also measured empathy or its sub-constructs using quantitative measures. These studies found positive main effects of social biofeedback on empathy [3], self-reported affect [31, 63], and usage of affect-related terminology [67, 110]. While excluded from our formal qualitative analysis, it must be noted that many quantitative studies in the review measured empathy or its sub-constructs. These found that social biofeedback leads to increased emotional perspective-taking [81, 136] and self-reported empathy [60, 61, 82, 95]. However, both positive [136] and null results have been reported for emotional convergence. These findings were not formally meta-analyzed due to the diverse range of empathy sub-constructs and questionnaires (including custom ones) employed.

### 4.3.4  Theme 4: Compassion and caregiving

Compassion is the feeling that arises when another person is suffering, which motivates helping behavior [47]. It differs from empathy in that it involves active support and not necessarily feeling the same emotion [13].

The most explicit examples of compassion were found in behavioral manifestations of helping behavior after seeing the physiological response of another user. Several studies found that users actively tried to soothe or calm down others [52, 65, 83, 110, 130] and sometimes censored their conversations in accordance with another person's physiological display [37, 110, 112]. In some studies, users reported consistently checking up on another's physiological display [52, 75, 109] and directly helping their co-user improve their performance on a task [46, 65, 130]. Altruistic behaviors were even noted in competitive settings [45, 73, 94], as well as expressions of sympathy and concern in adversarial contexts [89, 90].

Studies also found that users made conscious efforts to show thoughtfulness and respond to the emotions of others to make them feel validated [67, 77, 98, 110]. Specifically, in studies describing asymmetrical social biofeedback systems, compassion was found in that secondary users were eager to show sympathetic support for primary users going through hardship such as physical exertion [26, 64, 88, 130] and effortful gameplay [109]. Finally, an intriguing way of showing compassion was through humor; indeed, some studies found that users initiated humor to dissipate feelings of social embarrassment when one person felt overly-exposed [32, 54, 64]. Moreover, in one quantitative study of prosocial behavior (charitable donating), no effect of social biofeedback was found [81].

### 4.3.5  Theme 5: Relationship skills for authentic social connection

Relationship skills describe the actions taken to establish and maintain positive, healthy, and rewarding relationships [56], ultimately fostering emotional experiences of belonging. These are separate from empathetic responses because they do not necessarily entail feeling or fully understanding another person's emotions and are also separate from exercising compassion in that they do not require an individual to be suffering. Our analysis showed that social biofeedback systems fostered relationship skills in two ways: by encouraging users to initiate new connections and maintain positive intimacy in existing relationships.

Shared displays of physiological data were used to naturally spark new conversations between people [32, 52, 64, 72, 75, 80, 98, 130] and support novel and spontaneous interactions with strangers [37, 64, 72, 88, 109]. Some studies reported particularly meaningful and spiritually rewarding interactions with strangers or acquaintances through shared displays [33, 55, 93, 122]. With regards to existing relationships, simply having access to the physiological activity of another person without explicit affiliative action seemed to enhance feelings of social presence [3, 27, 55, 65, 119, 125]. However, users went further to maintain connections by initiating spontaneous play, such as playfully bothering others [32, 40, 46, 64, 65, 95, 120] and naturally collaborating without it being required in a task [2, 31, 72, 73, 130]. Studies also found that opportunities for unfiltered emotion sharing were welcomed to practice reciprocal trust and honesty, leading to enhanced emotional closeness [54, 77, 116, 133].

Interestingly, some studies also found a heightened sense of responsibility for others [63, 88, 120]. However, some studies found that social biofeedback may instead promote impersonal connections, by reducing necessary effort to understand others and distracting from authentic interactions [77, 78]. Likewise, intimate awareness and connectedness were also reported when viewing social biofeedback without explicit effort or action made by either senders or receivers [91, 133].

### 4.3.6 Theme 6: Motivation, performance, and coordinative effort

A recurrent finding in the reviewed studies was that interacting with social biofeedback systems motivated task-based persistence. In some studies, feeling observed by others motivated more effort to perform [26, 88, 123] and created opportunities for shared goal setting [93, 130]. Social biofeedback also naturally created a sense of healthy competition between athletes [94, 121] and even created feelings of competition when there was no task at all. In those studies, the controlling of physiological responses itself became a competitive endeavor [3, 64]. In competitive contexts, social biofeedback was used to strategically enhance chances of better performance [32, 40]. However, being intimately aware of others' physical capabilities also created inopportune pressure to perform, leading to giving up [85, 93, 121] and risks of over-exertion due to harmful competition [85, 130]. Despite this, high levels of engagement and fun were reported across many studies where users were engaged in tasks [31, 32, 40, 46, 64, 72, 87, 88, 95, 107, 130].

Coordinative effort was also an essential aspect in social biofeedback systems embedded in task-specific contexts. Studies found that users took the initiative to find creative ways to interact with a co-user to achieve a common goal, including developing new norms, rules and vocabularies being socially constructed around shared displays [2, 65, 95, 107, 130]. Teamwork also became prominent in asymmetrical systems where users depended on each other for feedback about their physiology [63, 95, 130]. Other studies found that spontaneous leadership initiatives emerged to maximize the efficiency of coordinated efforts, although this led to both positive [107] and negative [142] outcomes. The presence of social biofeedback also encouraged users to put more effort into communicating with other users [31, 46, 63, 72, 73, 88, 98, 130].

## 4.4 Meta-analysis

The meta-analysis for synthesizing the effectiveness of social biofeedback interactions on positive emotions included six papers involving 204 participants (30.5 years ± 6.42) (see Appendix E for study characteristics). Outcomes of assessments measuring positive affect were synthesized (e.g., Positive and Negative Affect Schedule (PANAS) [132], Networked Minds Social Presence (NMSP) [5], Self-Assessment Mannikin (SAM) [8]). The main results of the meta-analysis demonstrate effect sizes of biofeedback from a partner on positive affect measurements that vary from small ($d$=0 and $d$=0.146 for audio-haptic biofeedback on SAM [30] and $d$=0.01, $d$=0,02 and $d$=0.19 for audio-haptic biofeedback on PANAS [81]) to medium ($d$=0.62 for audio, visual, and audio-visual biofeedback on SAM [63]; 0.62 for visual biofeedback on PANAS [5]) to large ($d$=1.34, $d$=1.38, $d$=1.46 for visual biofeedback on the NMSP measurement [113]) (see Appendix F for details)**.** The most effective study, which measured positive affect using the NMSP measurement [113], was the only study to report a significant effect of feedback from three conditions: EEG ($d$=1.46; 95% CI=[0.78, 2.11]), respiration ($d$=1.38; 95% CI=[0.71, 2.04]), and EEG and respiration combined ($d$=1.34; 95% CI=[0.68, 2.00]). The other five studies reported effects of social HR biofeedback on self-reports of positive affect; however, none demonstrated significant effects.

## 5 DISCUSSION

Despite methodological advances in psychophysiological data analysis and biosensing technology, our understanding of physiological computing as an interaction paradigm remains fragmented. Existing research focusing on social uses of biofeedback has suggested that physiological data sharing constitutes a communication medium often associated with positive user experiences. However, it is unclear what communication contexts are associated with prosocial outcomes, as well as whether social biofeedback can promote more lasting interpersonal effects. In this paper, we systematically reviewed empirical progress in social applications of biofeedback over the last two decades. We identified physio-temporal and social contextual characteristics surrounding biofeedback-mediated communication, and highlighted socio-emotional competences associated with physiological data sharing. We also developed the Social Biofeedback Interactions framework to articulate the current physiological-social space based on the literature we reviewed. In the following section, we discuss the implications of our synthesis and propose opportunities for research and design with social biofeedback.

### 5.1 What types of interactions occur in physiological-social space, and with whom?

#### 5.1.1 Physio-temporal context of social biofeedback

As new communication technologies expand humans' reach in time and space [57], we suggest this is no different for social biofeedback systems. Using the Time-Space taxonomy of groupware [34], we found that current social biofeedback systems are designed for use more in synchronous time than asynchronous ways. The idea of synchronicity is generally important in computer-mediated communication because it provides an understanding of user context [23]. We found synchronicity was important for overcoming physical distance while performing social activities involving both competition and collaboration, including games like Space Connection [95] and FitBirds [85] and activities from meditation to distributed running. In these situations, biofeedback cues were used to inform co-users of each other's real-time states and as a mechanic to drive the social task itself. Interestingly, we also found this was the case for synchronous and co-located social biofeedback systems, e.g., [2, 32, 46, 54, 64, 87] demonstrating the informational role of social biofeedback in synchronous interactions.

However, synchronicity is not always practical over distance and may create unwanted pressure to constantly share biofeedback and undue responsibility for others' wellbeing when used for continuous, direct communication. Asynchronicity bypasses these drawbacks with two major types of interactions: awareness systems and instant messaging. First, following Weiser's notion of calm computing, we found awareness systems were all conceived to minimize attentional requirements (e.g., BioCrystal [112], ExternalEyes [37], Wigglears [98], Open Heart Helmet [130] and MoodLight [120]). With these systems becoming increasingly ambulatory, asynchronous co-located systems encouraged interactions that were not necessarily always focused on the display itself. For instant messaging, users also felt an increased connection to each other after using social biofeedback-enhanced systems (e.g., Significant Otter [77], HeartChat [52] and Animo [75]). We suggest that asynchronous systems create subliminal nudges like reminding others to be mindful of each other's emotional states. This supports the ideas of 'ambient co-presence' in polymedia [86], and that asynchrony in communication is more useful to stay in touch than to relate specific information about a person's wellbeing [57].

### 5.1.2 The Social Biofeedback Interactions framework

To articulate the current physiological-social interaction space, we formulated the Social Biofeedback Interactions framework based on [16]'s directions for social biofeedback and concurrent ideas of symmetry in media space [129]. This framework allows us to make informed observations about the different implications of social biofeedback system designs, as shown in Table 2.

Table 2: Table showing biofeedback parameters and implications for both symmetrical and asymmetrical social biofeedback systems

| System classification | Biofeedback parameter | System applications and implications |
|---|---|---|
| Asymmetrical | Biofeedback access granted to primary and secondary users | The main purpose of using this type of asymmetrical social biofeedback set-up is for celebrating self-expression of emotional state while raising awareness for bystanders in the surrounding environment. As emotional self-expression through biofeedback can be daunting, granting access to the primary user can alleviate self-presentation concerns as well as promote introspection. |
| | Biofeedback access granted only to secondary user | Intentionally designing the primary user out of the social biofeedback loop necessitates the acceptance of a loss of agency associated with self-disclosure, and trust. For those reasons, the most common application of this type of system to date is in the laboratory to conduct experiments of social perception of physiological signals. However, this type of asymmetricity has been useful in sporting events, where physiology reflects exertion effort as opposed to more intimate emotions. |
| Symmetrical | Biofeedback contains data from both users | Seeing another user's biofeedback as well as one's own has mostly been implemented in games and sport to engender competition and drive individual performance to rival that of a co-user's. This system design is also useful for mobile messaging applications, in a similar way to text-based texting where both users can see their own and their interlocuter's responses. Data from both users can also be presented in aggregated form, which can dispel concerns about social image, create opportunities for social play and for collaborative efforts. |
| | Biofeedback contains only data from the other user | This type of design enables users to receive physiological information of one another while not being explicitly aware of what physiological data they themselves are transmitting. The most common application for this was to increase intimate connectedness between two users in a way resembling natural communication. |

We used this framework to identify patterns of relationship composition in symmetrical and asymmetrical social biofeedback systems. Perhaps unsurprisingly, symmetrical systems had the most potential to connect close relations (family, friends) and romantic partners. This was the case both when users could see their own biofeedback and when they only had access to that of their co-users. This finding is in line with research showing that interpersonal closeness is fostered through reciprocity in computer-mediated communication [62]. However, adding onto [38]'s findings, we also found that symmetry is not always necessary to facilitate dyadic closeness; simply perceiving the biofeedback of another can increase empathy (e.g., [136]) and bring together people who do not necessarily know each other. We suggest two possible explanations: physiological signals are inherently intimate [38] and volitional act of sharing itself is perceived as an act of openness, encouraging secondary users to feel (and potentially become) closer. Overall, our findings align with previous research showing that different communication channels are used in different interpersonal relationships to achieve varying levels of electronic intimacy [74].

Our analysis also revealed that symmetrical systems were more easily embedded into closed group settings but less easily integrated into open public contexts than asymmetrical systems. In the group settings, symmetrical social biofeedback was used to unite people sharing the same hobbies, from competitive gaming to sport. Instead, where biofeedback was asymmetrical in groups, this occurred where a leader was established, suggesting that social biofeedback can be used for teaching purposes or when one group member is more dominant than the others. This follows trends of biofeedback sharing in assistive contexts (e.g., [126]). In addition to fostering connection [119], we suggest that these interactions enhance the experience of a social task, instead of the outcomes of social tasks. Depending on the task, this can be done using competition or collaboration, with symmetrical and asymmetrical biofeedback systems, respectively.

Finally, in the public realm, we found that asymmetrical systems could be used to provoke interactions between strangers or acquaintances (e.g., colleagues). In many of these cases, biofeedback displays were used as conversation starters, and social biofeedback functioned as a means of connection. Asymmetrical systems were also used in mass communication to connect supporters with athletes and fans with online influencers. On the other hand, symmetrical systems integrated into public settings were used to enhance audiences' social engagement.

## 5.2 What kind of socio-emotional competences can be practiced and developed during interactions through social biofeedback systems?

In this section, we discuss our qualitative analysis of socio-emotional competences to reflect on the support implications of emergent themes and to answer the question: as a new generation of physiological computing systems, what benefits can social biofeedback systems bring?

### 5.2.1 Self-oriented competences

The first two themes we identified were (1) mindful self-awareness and (2) self-reflection and regulation of affective states, both related to intrapersonal social-emotional competences. First, as an emerging topic in HCI, *state* mindfulness is an important skill to cultivate as it is positively correlated with wellbeing independently of *trait* mindfulness [11]. We found that receiving feedback about one's own physiology was associated with heightened self-awareness, detachment, and the perception of new relationships. This is consistent with existing research showing that components of mindfulness can be fostered using individual-basis biofeedback techniques [49]. We propose that social biofeedback systems also have this potential, despite being embedded in social communication pipelines and social tasks. In addition, receiving biofeedback from others was associated with calmness, meditative states and attitudinal shifts towards openness and acceptance of others, implying that sharing biofeedback with others promotes certain facets of state mindfulness, compared to merely receiving one's own. However, due to the cross-sectional nature of our results, we cannot make inferences about mindfulness as a trait.

Our review found that social biofeedback systems helped users reflect in elaborate ways about their emotional states and actively attempt to regulate them (theme 2). Self-reflection took the form of active questioning of biofeedback displays, and both internal and external efforts to understand and evaluate the causes of emotional arousal. Sometimes this judgment may not always be positive, e.g., [93], and we suggest this could lead to amplified stereotype threat. Self-reflection also occurred spontaneously and without direct instruction, implying that while many attempts at improving self-reflection are embedded in interventions using preventative design (e.g., cognitive-behavioral therapy [104]), self-reflection can also be promoted using active design approaches involving social biofeedback [13]. The findings from our meta-analysis further suggest that the presence of these social biofeedback systems could help support positive emotions. In this sense, we propose that social biofeedback systems hold immense promise for furthering the Positive Computing agenda [13].

Finally, social biofeedback systems enhanced self-regulation. This is an expected outcome considering self-regulation is the primary goal of biofeedback [53, 139]. However, a key finding was that secondary users also regulated their behavior when they were aware that the biofeedback of another was on display. We suggest this could be a mechanism of empathy, which is a component of emotional intelligence like self-regulation [7].

### 5.2.2 Others-oriented competences

We found three main themes related to others-oriented social-emotional skills: (1) empathy, (2) compassion and caregiving, and (3) relationship skills for authentic social connection. Empathy in itself was the target of a few studies, e.g., [27, 52, 61, 81, 82, 95]. It is a particularly relevant construct as nonverbal communication cues are critical in empathy development [50], and thus their absence in technology-mediated communication constitutes a significant barrier to human-human interaction [13]. Our work suggests that social biofeedback can address this; indeed, as demonstrated in our thematic and meta-analyses, sharing biofeedback was associated with both affective (e.g., emotional convergence and physiological synchrony) and cognitive forms of empathy (e.g., mind perception, mentalizing). Importantly, this did not only occur during direct communication but also during shared social activities. In line with research on 'motivated empathy' [27, 141], we suggest that experiencing empathy through social biofeedback may further motivate altruism and cooperative behaviors. Our findings also support the claim by [13] that adding digital medium-specific strategies, in this case, social biofeedback, is a viable design strategy that should be adopted to enhance technologically mediated empathy.

Another significant theme in our analysis was compassion and caregiving, which we found were exhibited spontaneously when caregiving was not required as part of a social task, and even more surprisingly, in competitive contexts. Significantly, compassion was not only shown towards known others but also exhibited during interactions with strangers. We suggest that the intimate nature of biosignals is likely to enhance sensitivity to others' suffering, which aligns with research showing that the human affiliative motivational system facilitates compassion with others [66]. We also put forward another explanation: because physiological signals may be more understandable when they represent high arousal (e.g., the meaning of a high HR is more intuitive than low HR), it could be that social biofeedback inherently helps make users aware of states of suffering (e.g., stressed [20]), more so than positive states. In that way, we propose that the informational value of physiological signals could be a motivator of compassion. Overall, we found that presenting social biofeedback can increase compassionate behavior among primary and secondary users alike.

Finally, the last theme we found was related to general relationship skills in building authentic social connections. We found that social biofeedback enabled new connections to be made between strangers and acquaintances and encouraged new creative forms of interactions between close relations. When interacting in physiological-social space, users reciprocally felt compelled to communicate honestly and openly. This is important given that online disinhibition effects linked to anonymity and asynchronicity can negatively affect social connections [124]. Instead, we found that social biofeedback can promote authentic and emotional self-disclosure, which has a *benign disinhibition* effect [69] that improves relationship skills.

### 5.2.3 Task-oriented competences

Task-oriented competences linked with motivation, performance, and coordinative effort (theme 6) can be developed when social biofeedback is used for its informational value. Overall, we found that task motivation was enhanced in competitive and collaborative contexts, and tasks involving social biofeedback were associated with high levels of engagement. This is consistent with previous research in assistive contexts (e.g., [126]), suggesting that biofeedback may provide situational awareness as well as conversational grounding necessary to accomplish and enjoy social tasks [42, 126]. Interestingly, in asymmetrical systems, motivation and effort expenditure were also enhanced where users were simply observed by secondary users, the latter of which were not involved in the task. This demonstrates that sharing biofeedback can have social facilitation effects, where the presence of others can enhance performance [140].

Our findings extend previous research by showing that social biofeedback in symmetrical systems can also increase motivation and coordinative effort. Specifically, bidirectional sharing of biofeedback can create opportunities for interdependence and teamwork, with a shift in focus from the individual to the group. However, we found that this can lead to adverse outcomes due to social comparison mechanisms [137]. Indeed, since physiological response is an objective proxy of performance, making the biofeedback of two users visible affords direct comparison of the self with others. We found this unwanted competition can lead to discomfort, reduce self-perceived competence and create desires to give up. Moreover, social biofeedback not only increases emotional connections between people which in turn facilitates task engagement, but it also provides informational cues that help users work together to drive task performance.

### 5.3 Ethical considerations, Challenges and Research Opportunities

Despite the identified positive themes of social biofeedback, there is a need to pay attention to ethical issues and challenges to guide future work. Firstly, the intimate nature of biosignals can expose users to privacy concerns. This can become more complicated when a user feels lacking control over one's physiology and feel being exposed to others [38], which can be associated with anxiety, discomfort and undue embarrassment. We suggest such distress about self-image is amplified particularly in asymmetrical systems where primary users do not have biofeedback access, and in larger social groups (e.g., public and mass contexts) where biodata is shared with strangers. While aggregation or abstraction of users' data can help alleviate the issues (see Table 2), it is also of importance to carefully consider interpersonal closeness between users, symmetricity of sharers and observers, and experiential location in implementing social biofeedback [33]. Furthermore, providing the ability to control the timing of personal data transmission can help alleviate privacy concerns [41], when designing in how physiological information is communicated. For instance, visualizations that provide less revealing information generate fewer privacy concerns [79].

Sharing physiological data can also induce feelings of excessive responsibility or concern towards the health or psychological status of co-users [80]. This could lead to attachment and dependency issues such as constantly checking another's vital signs, especially when display interfaces are portable and ubiquitous. We suggest this poses a potential risk when biodata is used for connection as opposed to information, when it is shared continuously (e.g., throughout the day), or when sharing is synchronous. Here, it is essential to consider interaction contexts and modalities that can support embedding biofeedback into our social world in non-obtrusive, ambient ways [79]. On the flip side of attachment, sharing biodata could also create alienation and uncanny valley effects [3]. Future research could also focus on other facets of user context [9], for example how motivational states influence whether physiological data should be abstracted in terms of cognitive or derived states to represent their interactions in physiological-social space [33]. These could shed light on the mechanisms behind some negative effects of symmetrical biofeedback systems.

Another focus for future research is exploring how social norms are challenged by the advent of social biofeedback systems. For instance, these systems could be deemed inappropriate for use with superiors (e.g., one with authority over the other) as they could promote faking social cues to generate desired outcomes or manipulation of someone based on perceived affective outcomes [33]. Analysis of the impact of power dynamics between sharers and observers on behavior during usage of shared physiological user interfaces could help identify what design factors may facilitate or hinder such detrimental processes [33]. It would also be interesting to investigate what factors could reduce potential risks of social biofeedback engendering unhealthy exertion in a certain competitive context such as fitness competitions, as individuals might strive to maintain a positive self-image to others. Also, further longitudinal and quantitative research is needed to ascertain whether interacting in physiological-social space leads to measurable increases in socio-emotional competences, perhaps as a factor of age. Longitudinal research would help to identify whether social biofeedback can help practice these skills or cause them to develop.

Finally, we suggest the proposed Social Biofeedback Interactions framework to serve as a basis for the design of future user interfaces of shared physiological data. Designers could use the framework to consider how and when user access or content of biofeedback should be granted or restricted, as well as to consider how new interactions could be formed between users, non-users, computers, and space. Furthermore, in line with the Positive Computing paradigm [13], we advocate for the use of dedicated and active design approaches in designing for mental and physical wellbeing with social biofeedback.

## 5.4 Limitations

There is room for improvement. Firstly, the main limitation arises from the diverse terms used in research on social biofeedback. During our initial search, some databases returned an excessive amount of entries which could not be downloadable due to some limited functionality of the digital libraries; thus, we limited our keywords to mitigate this, which could have led to the exclusion of relevant papers. For instance, by favoring umbrella terms, we may have excluded papers only specifying unimodal feedback (e.g., neurofeedback) and specific methods (e.g., electroencephalography). We also acknowledge that using broader social terminology (e.g., "public") could also have returned studies where social interplay was not explicitly considered. Secondly, the data investigated in our meta-analysis was limited due to the small number of articles given a diverse range of evaluation measures and methods used in the field, often lacking quantitative approaches. We encourage future research to use standardized, replicable methods and explore quantifiable constructs like empathy.

## 6 CONCLUSION

This systematic review is one of the first to consider shared user interfaces of physiological data, and the first to explore their contextual characteristics and enduring effects on social-emotional competences. We identified 64 key articles and synthesized characteristics of social biofeedback interfaces and contexts and created the Social Biofeedback Interactions framework that synthesizes the affordances of such systems based on biofeedback access and content. We learned that synchronicity lends itself to the use of social biofeedback as an informational cue, while over physical distance, biofeedback is used for connection. We also found that relationship composition plays a role in how social biofeedback is used. Finally, we found that social biofeedback can foster social-emotional competences on different levels: intrapersonal (mindful self-awareness and self-reflection and regulation), interpersonal (empathy, compassion and caregiving and relationship skills), and in relation to tasks (motivation, performance, and coordinative effort). Promisingly, our paper demonstrates the potential for social biofeedback to augment current technologies supporting human-human communication. Future work is needed to consider the role of social norms in adopting social biofeedback as a communication medium – not only their influences on interaction quality, but also how norms themselves are bound to change.

### ACKNOWLEDGMENTS

We thank the anonymous reviewers for their insightful feedback and suggestions. Youngjun Cho was supported in part by the Foreign, Commonwealth & Development Office through the AT2030 Programme (www.AT2030.org).

**Shared User Interfaces of Physiological Data: Systematic Review of Social Biofeedback Systems and Contexts in HCI**

Clara Moge, Katherine Wang, and Youngjun Cho

**APPENDICES**

**A. Flowchart of Our Methodological Processes Undertaken for This Systematic Review and Meta-Analysis Guided by PRISMA**

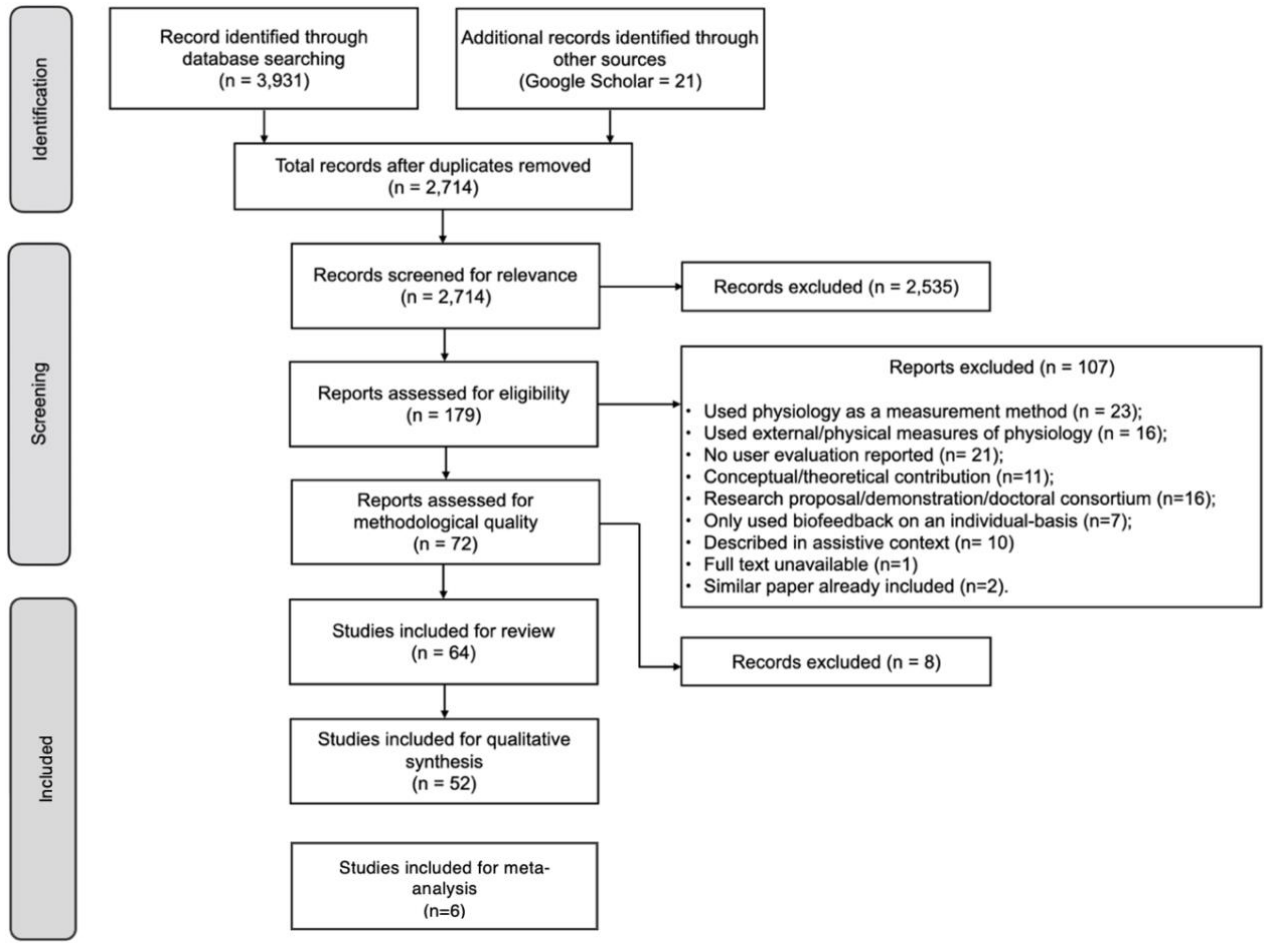

**Shared User Interfaces of Physiological Data: Systematic Review of Social Biofeedback Systems and Contexts in HCI**

Clara Moge, Katherine Wang, and Youngjun Cho

**B. Publication Year and Frequency of Reviewed Studies**

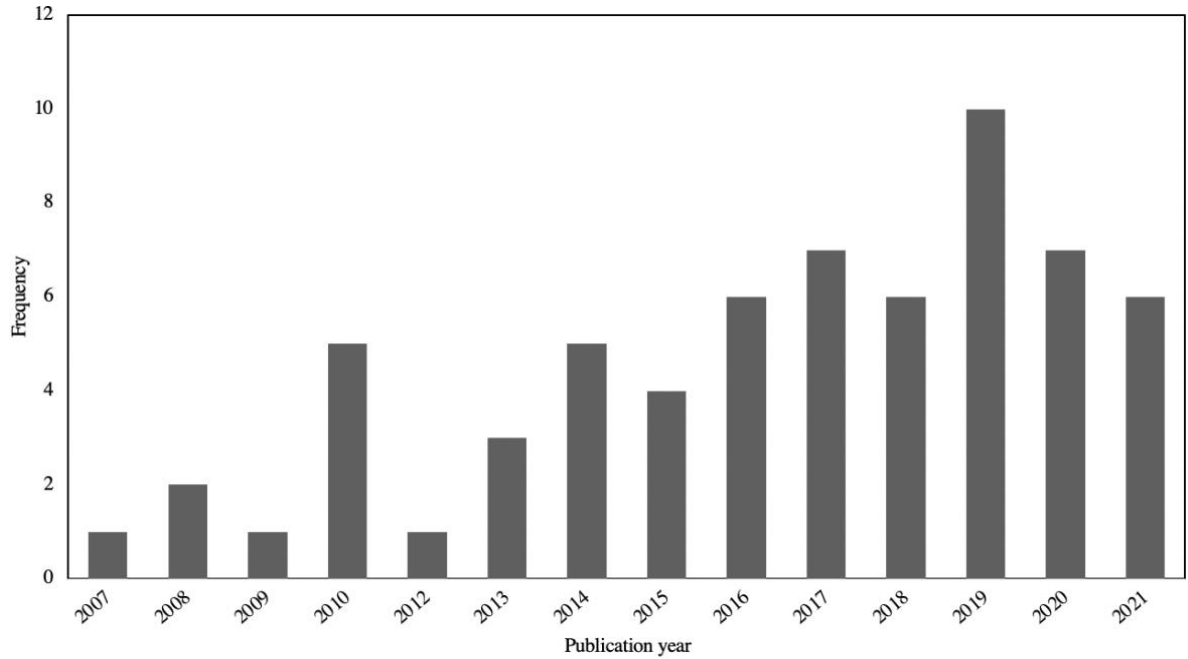

**C. General Characteristics of Methodologies for Reviewed Studies**

| Study | Social biofeedback use context(s) | Study design | | Environ-ment | | Evaluated measures | | | | Evaluation methods | | | | |
|---|---|---|---|---|---|---|---|---|---|---|---|---|---|---|
| | | Qualitative | Quantitative | Lab | Field | Physiology | Psych. | Performance / behavior | UX | Questionnaire | Interview | Observation | Diary/ daily notes | Data logs |
| Al Mahmud et al. (2007) | Social play | X | - | - | X | - | - | - | X | - | X | X | - | - |
| Aslan et al. (2020) | Tele-social communication | X | X | - | X | - | - | - | X | X | X | - | - | - |
| Chanel et al. (2010) | Public interactive performances | - | X | X | - | - | X | - | - | X | - | - | - | - |
| Curmi et al. (2013) | Online content sharing | X | - | - | X | - | - | - | X | - | X | X | X | X |
| Curran et al. (2019) | Online content sharing | X | X | X | - | - | X | X | - | X | X | - | - | - |
| D'Souza et al. (2018) | Social play | X | - | X | - | - | X | X | X | - | X | X | - | X |
| Dey et al. (2017) | Social play | X | X | X | - | X | X | - | - | X | X | - | - | - |
| Dey et al. (2018) | Social play | - | X | X | - | X | X | - | - | X | - | - | - | - |

**Shared User Interfaces of Physiological Data: Systematic Review of Social Biofeedback Systems and Contexts in HCI**

Clara Moge, Katherine Wang, and Youngjun Cho

| Study | Context | | | | | | | | | | | | | |
|---|---|---|---|---|---|---|---|---|---|---|---|---|---|---|
| Dey et al. (2019) | Social play | - | X | X | - | X | X | - | X | X | - | - | - | X |
| Elagroudy et al. (2008) | Public interactive installations | X | - | X | - | - | - | - | X | X | - | - | - | - |
| Fajardo et al. (2008) | Public interactive wearables | X | - | - | X | - | - | - | X | X | - | X | X | - |
| Frey (2016) | Social play | X | X | X | - | - | X | - | X | X | - | X | - | - |
| Frey et al. (2018) | Tele-social communication | X | X | X | - | X | X | X | X | X | X | - | - | X |
| George & Hassib (2019) | Social play | X | X | X | - | X | - | X | X | X | - | - | - | X |
| Gervais et al. (2017) | Multi-user relaxation | X | - | X | - | - | - | - | X | X | X | - | - | - |
| Hassib et al. (2017) | Tele-social communication | X | - | - | X | - | X | - | X | X | - | - | - | X |
| Howell et al. (2016) | Public interactive wearables | X | - | - | X | - | - | - | X | X | X | X | - | - |
| Howell et al. (2019) | Public interactive installations | X | - | X | - | - | - | - | X | - | X | - | - | - |
| Janssen et al. (2010) | Tele-social communication | - | X | X | - | - | X | X | - | X | - | - | - | X |
| Järvelä et al. (2021) | Social meditation | - | X | X | - | X | X | - | - | X | - | - | - | X |
| Järvelä et al. (2019) | Social meditation | - | X | X | - | X | X | - | - | X | - | - | - | X |
| Karaosmanoglu et al. (2021) | Social play | X | X | X | - | X | X | - | X | X | - | - | - | X |
| Khot et al. (2015) | Public interactive installations | X | - | - | X | - | - | - | X | - | X | X | - | - |
| Kim et al. (2015) | Tele-social communication | X | - | X | - | - | - | - | X | - | X | X | - | - |
| Kuber & Wright (2013) | Tele-social communication | X | X | X | - | - | X | - | X | X | X | - | - | X |
| Lee et al. (2014) | Tele-social communication | - | X | X | - | - | X | X | - | X | - | - | - | X |
| Li et al. (2018) | Social play | X | - | - | X | X | - | - | X | - | X | - | - | X |
| Li et al. (2019) | Social play | X | - | - | X | - | - | - | X | - | X | - | - | - |
| Liu et al. (2017a) | Tele-social communication | X | - | - | X | X | X | - | X | X | X | - | - | X |
| Liu et al. (2017b) | Mediated social interaction | X | X | X | - | - | X | - | X | X | - | - | - | - |
| Liu et al. (2019) | Tele-social communication | X | - | - | X | X | - | - | X | X | X | - | X | X |
| Liu et al. (2019) | Mediated social interaction | - | X | X | - | - | X | X | X | X | - | - | - | X |
| Liu et al. (2021) | Tele-social communication | X | - | - | X | - | - | X | X | X | X | - | X | X |
| Ma et al. (2020) | Social exertion | X | - | X | - | - | - | - | X | - | X | - | - | - |
| Magielse et al. (2009) | Social play | X | - | - | X | - | - | X | X | X | X | X | - | - |
| Mauriello et al. (2014) | Social exertion | X | - | - | X | - | - | X | X | X | X | X | - | - |
| Merrill & Cheshire (2016) | Mediated social interaction | X | X | X | - | - | - | - | X | X | - | - | - | - |

**Shared User Interfaces of Physiological Data: Systematic Review of Social Biofeedback Systems and Contexts in HCI**

Clara Moge, Katherine Wang, and Youngjun Cho

| | | | | | | | | | | | | | | | |
|---|---|---|---|---|---|---|---|---|---|---|---|---|---|---|---|
| Merrill & Cheshire (2017) | Mediated social interaction | X | X | X | - | - | - | X | X | X | - | - | - | - | X |
| Min & Nam (2014) | Tele-social communication | X | - | X | - | - | - | - | X | - | X | - | - | - | - |
| Moran et al. (2016) | Spiritual practice | X | - | - | X | - | - | - | X | - | X | - | - | - | - |
| Mueller et al. (2010) | Social exertion | X | - | - | X | - | - | - | X | - | X | - | - | - | - |
| Muños et al. (2016) | Social play | X | X | X | - | - | - | - | X | X | X | - | - | - | - |
| Peng (2021) | Public interactive wearables | X | - | - | X | - | - | - | X | - | - | X | X | - | - |
| Perttula et al. (2010) | Public interactive performances | X | - | - | X | X | - | - | X | X | - | X | - | - | X |
| Robinson et al. (2020) | Social play | X | X | X | - | - | X | X | X | X | X | - | - | - | X |
| Robinson et al. (2017) | Online content sharing | X | X | X | - | - | - | X | X | X | X | - | - | - | X |
| Rojas et al. (2020) | Face to face communication | X | X | X | - | - | - | X | X | - | X | X | - | - | X |
| Roseway et al. (2015) | Public interactive installations | X | - | - | X | X | X | - | X | X | - | X | - | - | - |
| Salminen et al. (2018) | Social meditation | - | X | X | - | X | X | - | - | X | - | - | - | - | X |
| Salminen et al. (2019) | Social meditation | - | X | X | - | X | X | - | - | X | - | - | - | - | X |
| Semertzidis et al. (2013) | Mediated social interaction | X | X | - | X | - | X | - | X | - | X | - | X | X | - |
| Shirokura et al. (2013) | Online content sharing | X | - | X | - | X | - | - | X | - | X | - | - | - | X |
| Slovák et al. (2012) | Face to face communication | X | - | X | X | - | - | - | X | - | X | - | - | X | - |
| Snyder et al. (2015) | Mediated social interaction | X | - | X | - | - | - | X | X | - | X | X | - | - | - |
| Sonne & Jensen (2014) | Social exertion | X | - | - | X | - | - | - | X | - | X | - | - | - | - |
| Stepanova et al. (2020) | Public interactive installations | X | X | - | X | X | X | - | X | X | - | - | - | - | - |
| Sugawa et al. (2021) | Public interactive performances | X | - | - | X | - | - | - | X | X | - | - | - | - | - |
| Sun & Tomimatsu (2017) | Tele-social communication | X | X | X | - | X | - | - | X | X | - | - | - | - | X |
| Walmik et al. (2014) | Social exertion | X | - | - | X | - | - | - | X | - | X | - | - | - | - |
| Werner et al. (2008) | Tele-social communication | X | - | X | - | - | - | - | X | - | - | - | X | - | - |
| Willemse et al. (2018) | Tele-social communication | - | X | X | - | - | X | - | - | X | - | - | - | - | - |
| Winters et al. (2021) | Mediated social interaction | - | X | X | - | - | X | X | - | X | - | - | - | - | - |
| Xue et al. (2019) | Public interactive installations | X | X | X | - | - | - | - | X | X | X | - | - | - | - |
| Zangouei et al. (2010) | Social play | X | - | X | - | - | - | X | X | - | - | X | - | - | - |

**Shared User Interfaces of Physiological Data: Systematic Review of Social Biofeedback Systems and Contexts in HCI**

Clara Moge, Katherine Wang, and Youngjun Cho

### D. Characteristics of Physiological Inputs, Sensors and Displays

*Physiological Inputs and Sensors*

The physiological metrics measured in the social biofeedback systems in our review fall into six categories: cardiovascular (n=31, 48%), electrodermal (n=7, 11%), respiratory (n=6, 9%), neural (n=4, 6%), temperature (n=4, 6%) and generalized (n=2, 3%). 14 systems (22%) fed back combinations of these signals to users, with the most common being cardiovascular-EDA metrics (n=6) and respiration-neural activity metrics (n=5). Metrics ranged most widely for cardiovascular measures, with seven studies manipulating HR biofeedback by being keeping it constant [80, 88, 89, 133], creating authentic preconstructions [58, 134] and substituting it with pre-recordings [131]. With respiration biofeedback, 14 studies focused on breathing rate, with four including amplitude as a complementary metric. The synchronicity of breathing patterns between users was also a feature of seven biofeedback systems.

*Physiological Displays*

Table D.1 shows specific biofeedback forms. In the visual domain (n=41), numerical forms of biofeedback were all used to display values of heart rate. Most studies visualized individual HR as an absolute value, while one study displayed the average HR across a group of people [99]. On the other hand, graphical biofeedback was used to visualize relative activity or progression of a physiological metric over time. Line graphs were used to show fluctuations in raw skin conductance [27, 70, 108, 117], changes in brain activity [78] and heart rate [26, 40, 79, 80] over time. Bar graphs were also used in one instance [89] with a textual caption for context. When used alone, text was either used to describe [80] or categorize arousal objectively (e.g. as "elevated", [45, 88, 89]), or to provide an interpretation of psychophysiological state (e.g., "stressed", [37]).

Table D.1: Biofeedback modalities and forms of reviewed studies

| Biofeedback modality | Biofeedback form | Studies |
| --- | --- | --- |
| Audio | Sound signal | [86] |
| | Body sounds | [55, 58, 63, 93, 134] |
| | Ambient sounds | [33, 41, 118, 121] |
| Visual | Numerical | [17, 26, 52, 70, 73, 79, 87, 99, 118, 120, 129] |
| | Graphical | [26, 27, 40, 70, 78–80, 89, 108, 117] |
| | Textual | [37, 45, 80, 88, 89] |
| | Animation | [2, 17, 31, 37, 40, 45, 52, 67, 70, 73, 75, 78, 82, 94, 118, 122] |
| | Avatar characteristics | [45, 46, 67, 77, 84, 94, 106, 140] |
| | Scene characteristics | [33, 60, 61, 63, 81, 112, 117, 121] |
| | Lighting | [32, 41, 54, 64, 78, 92, 111, 119] |
| | Actuated movement | [92] |
| Haptic | Vibration | [3, 41, 90, 109, 117, 124, 131] |
| | Pressure | [90] |
| | Temperature | [71, 133] |
| Audio-visual | Sound, scene effects | [63, 115] |
| Audio-haptic | Sound, vibration | [29, 30] |
| Visual-haptic | Movement | [65, 70, 97] |

**Shared User Interfaces of Physiological Data: Systematic Review of Social Biofeedback Systems and Contexts in HCI**

Clara Moge, Katherine Wang, and Youngjun Cho

## E. Characteristics of Studies Included in the Meta-Analysis

Table E.1: Characteristics of studies included in the meta-analysis

| Study | N | Age (M±SD) | Feedback Condition | Feedback Type | Affective Assessment |
|---|---|---|---|---|---|
| Dey (2017) [31] | 26 | 39.5(5.2) | HR | Visualization | PANAS |
| Dey (2018) [29] | 18 | 30.9(6.8) | HR | Audio-haptic | PANAS SAM |
| Salminen (2018) [112] | 44 | 27(6.5) | EEG | Visualization | NMSP |
|  |  |  | Respiration |  | NMSP |
|  |  |  | EEG & respiration |  | NMSP |
| Dey (2019) [30] | 24 | 30.2(6.7) | HR (active task) | Audio-haptic | PANAS SAM |
|  |  |  | HR (passive task) |  | PANAS SAM |
| Liu (2019) [80] | 62 | 37.47(10.16) | HR | Visualization | PANAS |
| Karaosmanoglu (2021) [63] | 30 | 26.03(3.18) | HR | Audio/Visual/Audio-visual | SAM |

## F. Forest Plot of Cohen's *d* in Post-Intervention Affective Test Results Between Feedback and No Feedback Conditions

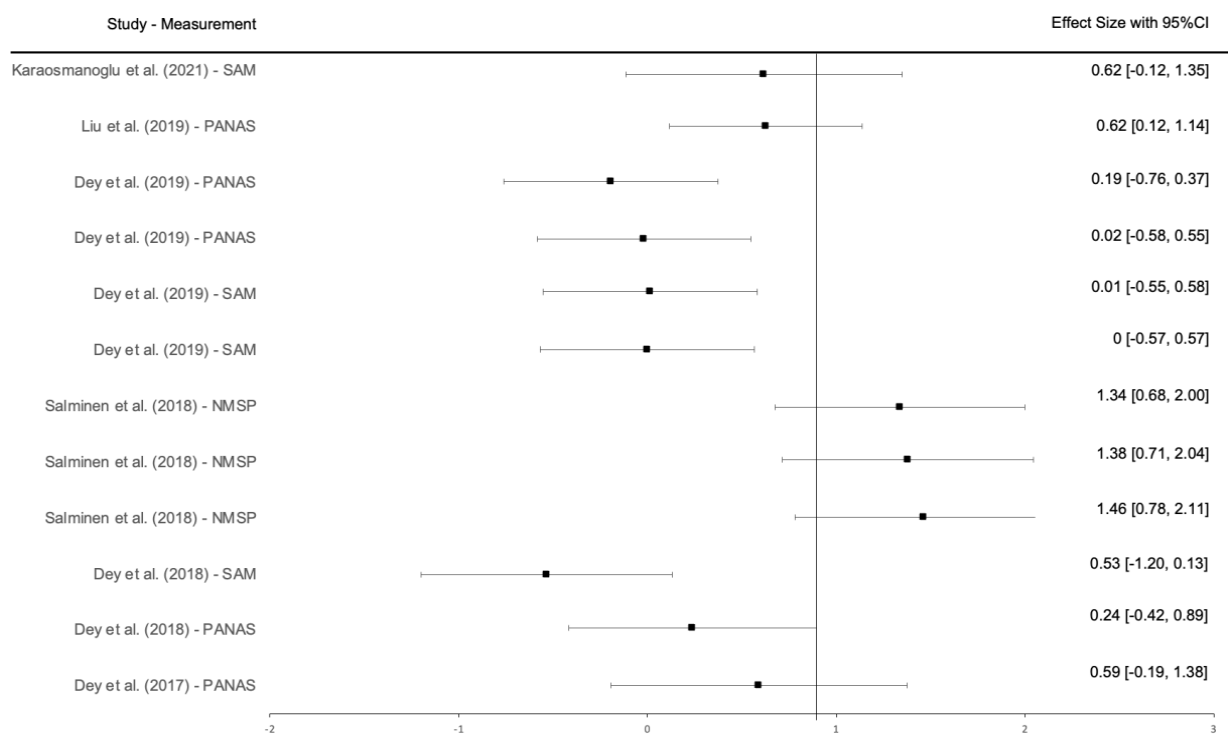

Figure F.1: This Forest plot depicts the main results of the meta-analysis. Effect sizes of post-intervention affective test results between feedback and no feedback conditions are indicated using Cohen's *d*. The left side favors control (no feedback) and the right side favors the presence of biofeedback (e.g., visual, audio, haptic) from a partner.